\documentclass[12pt,preprint]{iopart}
\usepackage{graphicx}
\usepackage{amssymb }

\newcommand {\ignore}[1]{}

\def\roughly#1{\mathrel{\raise.3ex\hbox{$#1$\kern-.75em
      \lower1ex\hbox{$\sim$}}}} \def\lsim{\roughly&lt;}
\def\gsim{\roughly&gt;}
\def\e6{E(6)}
\def\321{$SU(3)_{c}\otimes SU(2)_L \otimes U(1)$}
\def\10{SO(10)}

\def\422{SU(4) $\otimes$ SU(2) $\otimes$ SU(2)}

\def\Tr{\mbox{Tr}\,}

\def\hbar{\hspace{0pt}\raisebox{1pt}{$-$} \hspace{-7pt} h}

\def\5{\overline 5}

\usepackage[dvips]{color}
\definecolor{Black}{named}{Black}
\definecolor{Red}{named}{Red}
\definecolor{Blue}{named}{Blue}
\definecolor{Green}{named}{Green}
\definecolor{Brown}{named}{Brown}

\def\10{$SO(10)$}
\def\21{SU(2) $\otimes$ U(1) }

\def\422{$SU(4) \otimes SU(2) \otimes SU(2)$}
\def\321{SU(3) $\otimes$ SU(2) $\otimes$ U(1)}
\def\lsim{\raise0.3ex\hbox{$\;<$\kern-0.75em\raise-1.1ex\hbox{$\sim\;$}}}
\def\gsim{\raise0.3ex\hbox{$\;>$\kern-0.75em\raise-1.1ex\hbox{$\sim\;$}}}

\newcommand{\nn}{\nonumber}

\newcommand{\eV}{\mathrm{eV}}
\newcommand{\keV}{\mathrm{keV}}

\newcommand{\GeV}{\mathrm{GeV}}

\begin{document}


\submitto{JCAP}

\title[X-ray photons from late-decaying  majoron dark matter]{X-ray
  photons from late-decaying  majoron dark matter}
\date{\today
}
\newcommand{\AddrOXF}{Oxford Astrophysics, Denis Wilkinson 
Building, Keble Road, OX1 3RH, Oxford, UK }
\newcommand{\AddrAHEP}{%
  AHEP Group, Institut de F\'{\i}sica Corpuscular --
  C.S.I.C./Universitat de Val{\`e}ncia \\
  Edificio Institutos de Paterna, Apt 22085, E--46071 Valencia, Spain}
\newcommand{\AddrDCC}{
Dark Cosmology Centre, Niels Bohr Institute, University of Copenhagen,
Juliane Maries Vej 30, DK-2100 Copenhagen, Denmark}
\newcommand{\AddrINFN}{
Istituto Nazionale di Fisica Nucleare,
Via Enrico Fermi 40, 00044 
Frascati (Rome), Italy }
\author{Federica Bazzocchi$^1$, Massimiliano Lattanzi$^2$, Signe Riemer-S\o rensen$^3$ and Jos\'e W F Valle$^1$}
\address{$^1$ \AddrAHEP} 
\address{$^2$ \AddrOXF and \AddrINFN}
\address{$^3$ \AddrDCC}
\ead{fbazzo@ific.uv.es}
\ead{mxl@astro.ox.ac.uk}
\ead{signe@dark-cosmology.dk}
\ead{valle@ific.uv.es}

\pacs{14.60.Pq, 12.60.Jv, 14.80.Cp}

\begin{abstract}
\noindent  

An attractive way to generate neutrino masses as required to
account for current neutrino oscillation data involves the spontaneous
breaking of lepton number. 
The resulting majoron may pick up a mass due to gravity. If its mass
lies in the kilovolt scale, the majoron can play the role of
late-decaying Dark Matter (LDDM), decaying mainly to neutrinos. In
general the majoron has also a sub-dominant decay to two photons
leading to a mono-energetic emission line which can be used as a test
of the LDDM scenario. We compare expected photon emission rates with
observations in order to obtain model independent restrictions on the
relevant parameters.  We also illustrate the resulting sensitivities
within an explicit \textsl{seesaw} realisation, where the majoron
couples to photons due to the presence of a Higgs triplet.

\end{abstract}

\maketitle


\section{Introduction}
\label{sec:introduction}

While solar and atmospheric neutrino
  experiments~\cite{fukuda:1998mi,ahmad:2002jz,eguchi:2002dm} are
  confirmed by recent data from reactors~\cite{kl:2008ee} and
  accelerators~\cite{Collaboration:2007zza} indicating unambiguously
  that neutrinos oscillate and have mass~\cite{Maltoni:2004ei},
  current limits on the absolute neutrino mass scale,
\begin{equation}
  \label{eq:m1}
  m_\nu \lsim 1~\eV
\end{equation}
that follow from beta~\cite{Drexlin:2005zt} and double beta decay
studies~\cite{Avignone:2007fu}, together with cosmological observations
of the cosmic microwave background (CMB) and large scale
structure~\cite{Lesgourgues:2006nd} preclude neutrinos from playing a
{\sl direct} role as dark matter.

However, the {\sl mechanism} of neutrino mass generation may provide
the clue to the origin and nature of dark matter. The point is that it
is not unlikely that neutrinos get their mass through spontaneous
breaking of ungauged lepton
number~\cite{Chikashige:1980ui,schechter:1982cv}. In this case one
expects that, due to non-perturbative quantum gravity effects that
explicitly break global symmetries~\cite{Coleman:1988tj}, the
associated pseudoscalar Nambu-Goldstone boson - the majoron $J$ - will
pick up a mass, which we assume to be at the kilovolt
scale~\cite{Berezinsky:1993fm}.
The gauge singlet majorons resulting from the associated spontaneous
L--violation will decay, with a very small decay rate $\Gamma$, mainly
to neutrinos.
However, the smallness of neutrino masses (Eq.~(\ref{eq:m1})) implies
that its couplings to neutrinos $g_{J\nu\nu}$ are rather tiny and
hence its mean life is extremely long, typically longer than the age
of the Universe.
As a result such majorons can provide a substantial fraction,
possibly all, of the observed cosmological dark matter.

Here we show how the late-decaying majoron dark matter (LDDM) scenario,
and in particular the majoron couplings $g_{J\nu\nu}$ and $g_{J\gamma\gamma}$ to
neutrino and photons respectively, can be constrained by cosmological
and astrophysical observations.

The paper is organized as follows. In Sec.~\ref{sec:constraints} we
describe the basic cosmological constraints on the LDDM scenario, in
Sec.~\ref{sec:x-rays} we describe the ``indirect detection'' of the
LDDM scenario and determine the restrictions on the relevant
parameters that follow from the x-ray background and the emission from
dark matter dominated regions. In Sec.~\ref{sec:xrayCMB} we compare
the sensitivities of CMB and x-ray observations to the LDDM scenario,
stressing the importance of the parameter
$R=\Gamma_{J\gamma\gamma}/\Gamma_{J\nu\nu}$. Finally in
Sec.~\ref{sec:particle-physics} we discuss an explicit \textsl{seesaw}
model realisation of the LDDM scenario, where the majoron couples to
photons due to the presence of a Higgs triplet.

\section{Cosmological constraints}
\label{sec:constraints}

The LDDM hypothesis can be probed
through the study of the CMB anisotropy spectrum.
In fact, current observations, mainly of the Wilkinson
Microwave Anisotropy Probe (WMAP) lead to
important restrictions.
Indeed, the LDDM scenario has been explored in detail within a
modified $\Lambda$CDM cosmological model;
in particular, it has been shown that the CMB anisotropies can
be used to constrain the lifetime
$\tau_J\simeq\Gamma_{J\nu\nu}^{-1}$ and the present
density  $\Omega_{J}=\rho_J/\rho_c$ of the 
majoron~\cite{Lattanzi:2007ux}.

The reason is that the late decay of majorons to neutrinos would
produce too much power at large scales, through the late integrated
Sachs-Wolfe effect, thus spoiling the CMB anisotropy spectrum.
WMAP third year data \cite{Spergel:2006} can be used to constrain:
\begin{equation}
\Gamma_{J \nu\nu}<1.3\times10^{-19}\mathrm{sec}^{-1},
\label{eq:gammanu-limit}
\end{equation}
at 95\% CL~\cite{Lattanzi:2007ux}.

This result is independent of the exact value of the decaying dark
matter particle mass, and is quite general, in the sense that a
similar bound applies to all invisible decays of cold or warm dark matter
particles \cite{Berezinsky:1996pb,Ichiki:2004vi,Gong:2008gi}.

The CMB spectrum can also be used to constrain the majoron energy
density.  This can be translated to a limit on the majoron mass in a
model-dependent way. Given the majoron mass $m_J$ and lifetime
$\tau_J$, the present majoron density
parameter $\Omega_J$ can be written as:
\begin{equation}
\Omega_J h^2=\beta \frac{m_J}{1.25\,\keV}e^{-t_0/\tau_J},
\label{eq:omj}
\end{equation}
where $h$ is the dimensionless Hubble constant, $t_0$ is the present
age of the Universe, and the parameter $\beta$ encodes our ignorance
about the number density of majorons. The normalization in
Eq. \ref{eq:omj} is chosen such that $\beta=1$ if (i) the
majoron was in thermal equilibrium in the early Universe; and (ii), it
decoupled sufficiently early, when all the quantum degrees of
freedom in the standard model of fundamental interactions were
excited.

This simple picture can be changed if: (i) The majoron could not
thermalize before it decoupled from the other species, or (ii) The
entropy generated by the annihilation of some particle beyond the
standard model diluted the majoron abundance after its decoupling.

In any case it is reasonable to assume that the majoron decoupled at
$T\gtrsim 170\,\GeV$ since its couplings to all the other particles in
the standard model (SM) are tiny.
Using the WMAP third year data, the following constraint on $\Omega_Jh^2$ can be
obtained (95\% C.L.) assuming the dark matter to consist only of 
majorons~\cite{Lattanzi:2007ux}:
\begin{equation}
0.09 \leq \Omega_J h^2 \leq 0.13
\end{equation}

Since Eq. (\ref{eq:gammanu-limit})
implies $\tau_J \gg t_0$, the above constraint together with
Eq. (\ref{eq:omj}) gives:
\begin{equation}
0.12\, \keV < \beta m_J < 0.17\, \keV
\label{eq:mj-constraint}
\end{equation}

Our ignorance of the details of the majoron production
mechanism, namely the value of $\beta$, can always be used in order to
accommodate additional restrictions to the majoron mass $m_J$ 
coming from observations of the large scale structures.

 The limits quoted above apply to the invisible decay
  $J\to\nu\nu$.  There exist also the very interesting possibility to
  use the CMB polarization to directly constrain the radiative decay
  $J\to\gamma\gamma$. 
  This is because photons
  produced by dark matter decay can inject energy into the baryonic
  gas and thus affect its ionization history. This will ultimately
  lead to modifications of the CMB temperature-polarization (TE)
  cross-correlation and polarization auto-correlation (EE) power
  spectra. In Ref. \cite{Zhang:2007zzh}, WMAP third year data are used to obtain
  the following constraint for the radiative decay width $\Gamma_{\mathrm{rad}}$ 
  of long lived dark matter particles like the majoron:
\begin{equation}
\zeta\Gamma_{\mathrm{rad}} < 2.4 \times 10^{-25} \mathrm{sec}^{-1},
\end{equation}
where $\zeta$ is an ``efficiency'' factor describing the fraction of
the decay energy actually deposited in the baryon gas. This depends,
among other things, on the energy of the emitted photon. As a rule of
thumb, consider that for redshifts $10<z<1000$, when most of the
hydrogen is neutral, photons with energy in the range $13.6\,\eV$ (the
hydrogen ionization threshold) to approximately $1\,\keV$ will transfer most
of their energy to the baryon gas through photo-ionizations, so $\zeta
\sim 1$.  On the other hand, the Universe is transparent with respect
to the propagation of photons with $E\gg1\,\keV$, and in this energy
range one expects to have $\zeta \sim 0$ and then no significant upper
limit on $\Gamma_{\mathrm{rad}}$ can be obtained from CMB
polarization.

\section{X-ray analysis}
\label{sec:x-rays}

In a variety of neutrino mass generation models with spontaneous
violation of lepton number, majorons have an effective interaction
term with photons
\begin{equation}
  \label{eq:Jgg}
g_{J\gamma\gamma} J \epsilon^{\nu\mu \rho\sigma} F_{\nu\mu}F_{\rho\sigma}\,.
\end{equation}
Majorons in the $\keV$ range are therefore expected also to decay radiatively
into two photons of energy $E_\gamma\simeq m_J/2$, 
since the decay can be considered to a very good approximation as
happening in the dark matter rest frame. This leads to a mono-energetic 
emission line as a characteristic signal of our decaying dark matter model.

Such an emission line could be possibly be detected both in the
diffuse x-ray background and in the emission from dark matter
dominated regions.  We now consider the constraints coming from both
kinds of observations. 

In the following, when necessary, we will consider a LDDM scenario within a
$\Lambda$CDM cosmology with $\Omega_{\Lambda}=0.75$,
$\Omega_{DM}h^2=0.11$, $\Omega_{b}h^2=0.022$, and $h=0.72$,
corresponding to the best fit values of the CMB analysis in Ref.~\cite{Lattanzi:2007ux}.

\subsection{Diffuse x-ray background} \label{sec:diffuse}

Photons produced in late majoron decays will show up in the diffuse
x-ray background, if the Universe is transparent with respect to their
propagation.  This is indeed the case after the Universe has been
completely reionized ($z\lesssim 10$): photo-ionization is no more
effective in absorbing the photon energy, simply because there are no
more neutral hydrogen atoms to be ionized.

The flux $F(E)$ of decay radiation at the present time ($z=0$) is given by \cite{Chen:2003gz}:
\begin{equation}
F(E) = \frac{c}{4\pi}\left(\frac{E}{E_\gamma}\right)^3\left.\frac{N_\gamma\Gamma_{J\gamma\gamma}n_J(z)}{H(z)}\right|_{1+z = E_\gamma/E}
\end{equation}
where $N_\gamma = 2$ is the number of photons produced in each decay,
$n_J(z)$ is the number density of majorons at redshift $z$, $H(z)$ is
the Hubble parameter, and $E_\gamma = m_J/2$ is the energy of the
photons produced in the decay. This can differ from the energy $E$ at
which we are observing due to the cosmological redshift of photons.
In other words, when looking today at an energy $E<m_J/2$ we can still
expect some signal from photons emitted in the past with energy
$E_\gamma = m_J/2$ that have been red-shifted to lower energies.

However we know from the CMB that the majoron is very long lived,
so that we expect the decay spectrum to be dominated by very recent decays.
We model the spectrum as mono-energetic with $E=E_\gamma$ and a flux given by:
\begin{equation}
F(E_\gamma) = \frac{c}{4\pi}\frac{N_\gamma\Gamma_{J\gamma\gamma}n_{0,J}}{H_0},
\end{equation}
where the subscript $0$ denotes quantities evaluated at the present time.

This should be compared with the observed diffuse x-ray flux from
ASCA~\cite{Gendreau:1995} and 
HEAO-1 \cite{Gruber:1999yr}, operating
in the 0.4-7 keV and 3-500 keV ranges respectively.
The flux can be modeled as 
\cite{Kawasaki:1997ah} (units are sec$^{-1}$ cm$^{-2}$ sr$^{-1}$):
\begin{equation}
F_{obs}(E) = 
\left\{
\begin{array}{ll}
 8\displaystyle\left(\frac{E}{\keV}\right)^{-0.4}, & 1\,\keV<E<25\,\keV,\\[0.5cm]
 380\displaystyle\left(\frac{E}{\keV}\right)^{-1.6}, & 25\,\keV<E<350\,\keV,\\[0.5cm]
 2\displaystyle\left(\frac{E}{\keV}\right)^{-0.7}, & 350\,\keV<E<500\,\keV,
\end{array}
\right.
\end{equation}
Below 1 keV, the strong galactic emission must be carefully removed in
order to find the extragalactic signal \cite{Chen:1995jc}, and
consequently we do not extrapolate the above approximation to lower
energy for the purpose of the present analysis.

\begin{center}
\begin{figure}[ht]
\includegraphics[clip,width=0.6\linewidth]{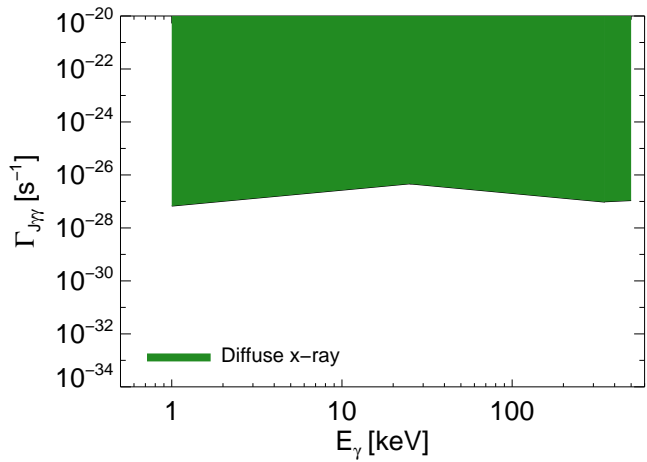}
\caption{Upper limit from the diffuse x-ray argument. The filled
  region is excluded. Data from \cite{Gendreau:1995,Gruber:1999yr}.}
\label{fig:decayrate0}
\end{figure}
\end{center}

Then, requiring $F(E_\gamma) \le F_{obs}$ yields an upper limit for
the majoron decay width to two photons $\Gamma_{J\gamma\gamma}$.  In
particular, in the range,
$1\,\keV\le E_\gamma\le 25\,\keV$ we have:
\begin{equation}
\frac{\Gamma_{J\gamma\gamma}}{\mathrm{sec}^{-1}}\lesssim 4.45\times 10^{-27}\left(\frac{h}{0.72}\right)\left(\frac{\Omega_Jh^2}{0.11}\right)^{-1}
\left(\frac{m_J}{\keV}\right)^{0.6}.
\end{equation}
This limit, together with the constraints at higher energies, is shown
in Fig.  \ref{fig:decayrate0}.

This simple analysis, and the resulting constraint, can be improved in
two ways. First of all, one can look for small distortions in the
smooth diffuse flux produced by a DM emission line that is possibly
lying well below the background signal. In addition, one can take into
account the contribution to the signal coming from the Milky Way. This
was applied to the HEAO-1 data in
Refs. \cite{Boyarsky:2005us,Boyarsky:2006fg}; in this way, the above
constraints can be improved by as much as three orders of magnitude
(see below). Finally, we note that bounds in the soft x-ray region can
be obtained from the observations of a high-resolution spectrometer
\cite{Boyarsky:2006hr}.

\subsection{X-rays from dark matter dominated regions} 
\label{sec:dmxray}

Observations of the x-ray emission from dark matter dominated regions
can be used to restrict the decay rate into photons and the mass of
any dark matter candidate with a radiative two-body decay. This
follows from the consideration that the detected flux from a dark
matter dominated object gives a very conservative upper limit on the
flux generated by dark matter decays in that object.

Since the dark matter in cosmological structures is practically at
rest (v/c $\approx 10^{-4}$), the line broadening due to motion of the
dark matter is negligible compared to the instrumental resolution of
current day x-ray detectors. Hence, a good instrumental spectral
resolution increases the sensitivity to a mono-energetic emission
line.

For majorons the 0.1-0.3~keV x-ray interval is very
interesting. Unfortunately this range is not accessible with any of
the standard CCD instruments on board the present x-ray observatories
{\it Chandra} and {\it XMM-Newton}. However, {\it Chandra} carries the
High Resolution Camera (HRC) which combined with the Low Energy
Transmission Grating (LETG) makes it possible to obtain spectra in the
0.07-10.0~keV range. The resolution of grating spectra is very high
($E_{FWHM}\approx 1$~eV) \cite{POG9} but all spatial information about
the photon is lost, except that it is known to origin from within the
field of view (apart from minor effects of scattering along the line
of sight).

To reach the maximum resolution requires bright point like sources
located at the aim point of the observation. Unfortunately dark matter
structures have spatially extended distributions and will produce a
very faint signal, if any. Extended sources can be thought of as made
up of many point sources, but then most of the sources are
off-axis. The effect of a source being off-axis is, that in
the detector plane, there is an ambiguity between angle and photon
energy, which gives a "smearing" towards lower energies and hence a
line broadening in the obtained spectra. The line smearing is energy
dependent and worst for high energies \cite{Riemer:2006b}.

No optimal sources for a search for dark matter decay line emission
have been observed with HRC/LETG. Still, from grating observations of
an active galaxy, we have improved the upper limit on the decay rate
from the dark matter halo in which the active galactic nuclei is
embedded by orders of magnitude. For photon energies above 0.3~keV
better constraints are obtained from conventional x-ray CCD
observations of merging clusters of galaxies such as the Bullet
Cluster \cite{Boyarsky:2006kc} or Abell~520 \cite{Riemer:2006b}.

\begin{figure}[ht]
\centering
	\includegraphics[clip,width=0.5\linewidth]{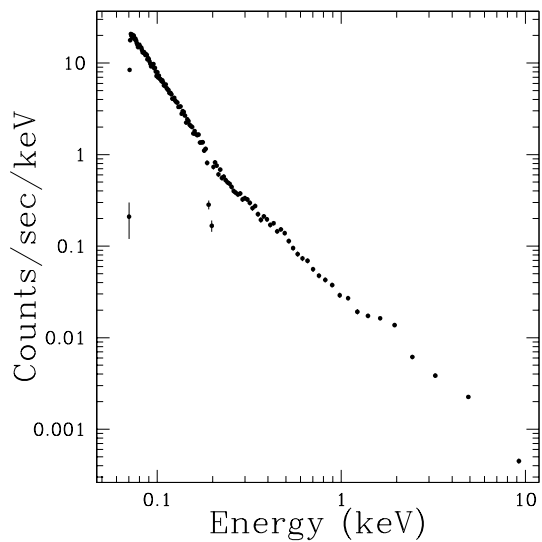}
	\caption{The observed {\it Chandra} HRC/LETG spectrum of
          NGC3227 (folded with the instrumental response).}
	\label{fig:spectrum}
\end{figure}

We have studied a {\it Chandra} HRC/LETG observation of the Seyfert 1
galaxy NGC 3227  (observation id 1591), shown in
  Fig. \ref{fig:spectrum}.  NGC3227 has a redshift of $z=0.004$,
which corresponds to a luminosity distance of $16.7$~Mpc. The data has been processed and
analysed with CIAO version 3.4 using CALDB version 3.4.0
\cite{Fruscione:2006}. The obtained spectrum is shown in
Fig. \ref{fig:spectrum}.

In general, the instrumental response cannot be unfolded from the spectrum in a
model independent way. Instead a model is folded with the instrumental
response and fitted to the data using $\chi^2$ statistics (here we have used the spectral fitting package {\it Sherpa} distributed with CIAO).

The model is used to determine an upper limit on the received
flux. Since no physical quantities are derived from the empirical
model, it is chosen to fit the data (and as such do not necessarily
represent a physical model of the emission). The data were split into
two intervals: 0.072-0.276~keV and 0.276-4.14~keV and fitted
separately to models composed of a power law and four Gaussians for
the lower interval and two power laws and two Gaussians for the higher
interval. In order to ensure that no emission lines are sticking above
the model, the fitted model was re-normalised so there were no bins in
the spectrum at more than $2\sigma$ above the model.

As mentioned above, any emission line is smeared out because of the
instrumental resolution. Towards lower energies the smearing has the
shape of a Gaussian with the width given by the instrumental
resolution. Towards higher energies, where the extension of the source
plays a role, the smearing depends on the overall distribution of the
dark matter. We have assumed an Navarro, Frenk \& White
(NFW)~\cite{Navarro:1996gj} profile for the dark matter halo of
NGC3227 with a scale radius of 15~kpc and a virial radius of
200~kpc. These are conservative representative values for galaxies, in
the sense that most galaxies have a smaller scale radius and a smaller
radius, leading to less smearing (and thereby tighter constraints).
The resolution is only sensitive to the full width at half maximum of
the density profile, so that choosing a different parameterization
(e.g.  Moore \cite{Moore:1999gc}) would not significantly alter the
results.

Since we are only interested in the upper limit on the measured flux,
it has been determined in slices of width $E_\gamma +
FWHM_{instrumental} > E > E_\gamma - FWHM_{smearing}$ instead of the
exact shape of the smeared lines (the difference between the two
methods is negligible \cite{Riemer:2006b}).

The mass of NGC3227 has been taken to be $10^{11}$ solar masses which
is conservatively low based on the luminosity of the galaxy
\cite{Bentz:2006qc}. The observational field of view is $\approx
25$~kpc at the distance of NGC3227, reducing the observed mass to about
a tenth of the total mass. This is probably an underestimate of the
observed mass, but a larger observed mass will only improve the
constraints.

Assuming only one kind of dark matter, the observed flux, $F_{obs}$,
at a given photon energy yields an upper limit on the decay rate from
two-body radiatively decaying dark matter:
\begin{equation} \label{eqn:maxdecay}
\Gamma _{J\gamma\gamma} \leq \frac{8 \pi F_{obs} D_L^2}{M_{fov}}  \, .
\end{equation}
The determined flux is dominated by the baryonic emission of the
galaxy, which varies with energy, introducing an apparent energy
dependence on the constraint.

\begin{center}
\begin{figure}[ht]
\includegraphics[clip,width=0.8\linewidth]{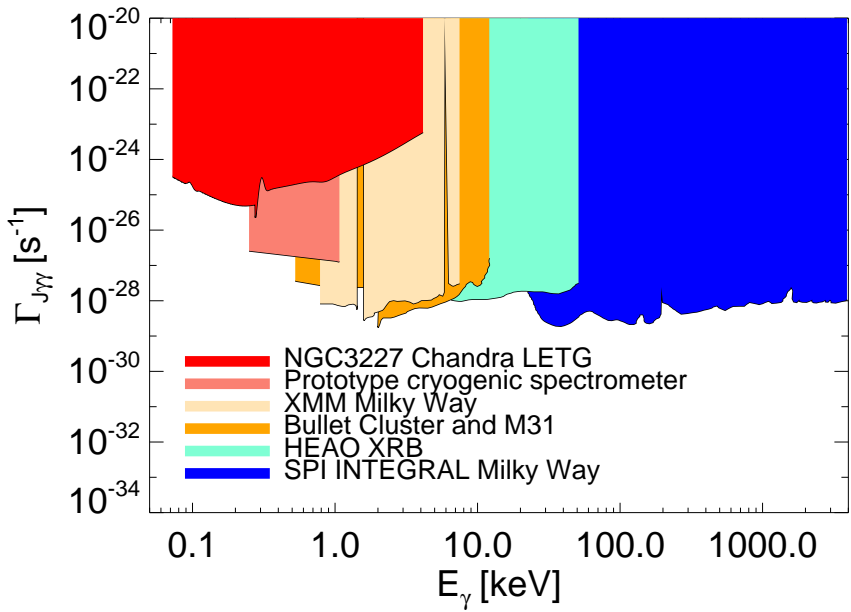}
\caption{ Upper limit on the decay rate from NGC3227 (red), the
    Milky Way halo observed with a prototype cryogenic spectrometer
    (salmon) \cite{Boyarsky:2006hr}, {\it XMM} observations of the
    Milky Way (sand) \cite{Boyarsky:2006ag}, {\it Chandra}
    observations of the Bullet Cluster \cite{Boyarsky:2006kc} and M31
    \cite{Watson:2006qb,Boyarsky:2007ay} (orange), HEAO-1 observations of the diffuse x-ray
    background (aquamarine) \cite{Boyarsky:2005us,Boyarsky:2006fg},
    INTEGRAL SPI line search in the Milky Way halo (blue)
    \cite{Yuksel:2007xh,Boyarsky:2007ge}. Filled regions are excluded.}
\label{fig:decayrate}
\end{figure}
\end{center}

The resulting constraint on the decay rate is shown in
Fig. \ref{fig:decayrate} together with earlier published
constraints.

\section{X-ray versus CMB} 
\label{sec:xrayCMB}

In the previous sections we have shown how the CMB can be used to
constrain the invisible decay $J\to\nu\nu$, while x-ray observations
can constrain the radiative decay $J\to\gamma\gamma$
\footnote{We do not consider the CMB polarization limit in the
    following because (i) it depends on the efficiency of the energy
    transfer to the baryonic gas and (ii) it turns out to be less
    constraining than x-rays in the regions of interest.}.  From a
theoretical point of view, a very important quantity is the branching
ratio of the decay into photons $BR(J\to\gamma\gamma)$ that, as long
as the decay to neutrinos is by far the dominant channel, is given by
the ratio of the decay widths:
\begin{equation}
BR(J\to\gamma\gamma) \simeq R=\frac{\Gamma_{J\gamma\gamma}}{\Gamma_{J\nu\nu}}.
\end{equation}

Note that the decay $J\to\nu\nu$ arises at the tree-level, while the
radiative $J\to\gamma\gamma$ mode proceeds only through a calculable
loop diagram. Beyond this, theory can not predict the expected value
of $R$, which is strongly model-dependent.
Here we use $R$ as a phenomenological parameter varying over the
wide range $10^{-25}-10^{-3}$ (see, for instance,
Fig.~\ref{fig:decayJgg.eps} below).  It should be clear that the
observations described in Secs.~\ref{sec:constraints} and
\ref{sec:x-rays} restrict $\Gamma_{J\nu\nu}$ and
$\Gamma_{J\gamma\gamma}$, leaving the branching ratio 
unconstrained.

However, we are also interested in knowing for which models the x-ray
observations can probe the decaying majoron dark matter hypothesis
with higher sensitivity than the CMB.  In particular, we expect that
models with large branching ratios will be better constrained by the
x-ray observations, since they will predict a larger production of
photons.

The x-ray limits presented in Sec.~\ref{sec:x-rays} have a mass
dependence, which we need to take into account in our assessment of
the relative constraining power of the two types of observations. We
know, however, from the CMB (see Eq. \ref{eq:mj-constraint}) that:
\begin{equation}
\frac{0.12\,\keV}{\beta}\le m_J\le \frac{0.17\,\keV}{\beta},
\end{equation}
at 95\% C.L. Fixing the value of $\beta$ is then equivalent
to fixing the majoron mass, apart from a small uncertainty (which we
take into account, see below). We express our
result in terms of $\beta$ instead than $m_J$.

\begin{center}
\begin{figure}[t]
\includegraphics[clip,width=0.8\linewidth]{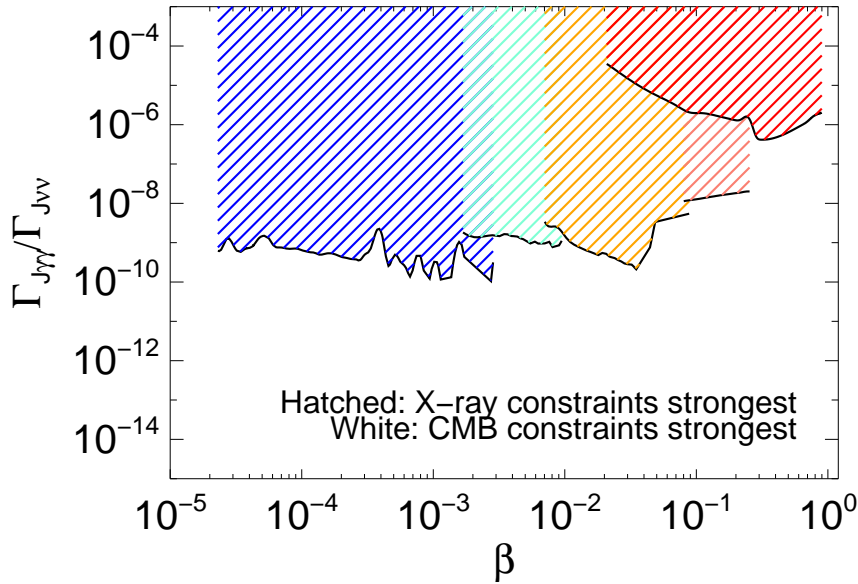}
\caption{Sensitivity of CMB and x-ray observations to the LDDM majoron
  scenario as a function of $\beta$ and
  $R=\Gamma_{J\gamma\gamma}/\Gamma_{J\nu\nu}$. The black lines
    are the loci of points where $R=R^*$, i.e., where the CMB and x-ray constraints
    (from a given object) are equivalent.  In the hatched regions above the lines, 
    X-ray constraints are stronger;
    below, CMB constraints are stronger. The color codes are the same
    as in Fig.~\ref{fig:decayrate}.
    See Sec.~\ref{sec:xrayCMB} for discussion.}
\label{fig:best-constraints}
\end{figure}
\end{center}
In order to compare the CMB and the x-ray constraints, we fix
  the value of $\beta$ and determine the corresponding observational
  ratio of $\Gamma_{J\gamma\gamma}/\Gamma_{J\nu\nu}$. According to the
  CMB constraints, we take the mass of the majoron to be equal to $m_J
  = 0.145\,\keV/\beta$, with an associated $1\sigma$ error of
  $\sigma_J=0.01\,\keV/\beta$. Then, we find the maximum
  $\Gamma_{J\gamma\gamma}$ allowed by the x-ray emission for this
  value of the mass (as explained in Sec. \ref{sec:dmxray}). The
  uncertainty in the exact value of the mass is taken into account by
  convolving the upper limits shown in Fig.~\ref{fig:decayrate} with a
  Gaussian of mean $m_J$ and variance equal to $\sigma^2_J$. Let us
  call this value $\Gamma_{J\gamma\gamma}^{\mathrm{max}}$.We also
  denote with $\Gamma_{J\nu\nu}^{\mathrm{max}}=1.3\times10^{-19}
  \mathrm{sec}^{-1}$ the CMB upper limit on the decay width to
  neutrinos. Then for the following value of the branching ratio:
\begin{equation}
  R^*=\frac{\Gamma_{J\gamma\gamma}^{\mathrm{max}}}{\Gamma_{J\nu\nu}^{\mathrm{max}}}
\end{equation}
the two sets of observations yield exactly the same constraining
power. In other words, for this particular value of the branching
ratio, it would be the same to constrain the decay rate to photons
using the x-rays and then obtain the decay rate to neutrinos using
$\Gamma_{J\nu\nu}=\Gamma_{J\gamma\gamma}/R$, or to do the contrary,
i.e. to use the CMB to constrain the invisible neutrino decay channel
and from that obtain a bound on the photon decay. Larger branching
ratios ($R>R^*$) will be better constrained by observations of the
x-ray emission, while smaller branching ratios ($R<R^*$) will be
better constrained by the CMB.
  
We repeated above procedure for $\beta$ ranging from $10^{-5}$ to
$1$, comparing the x-ray constraints of Fig.~\ref{fig:decayrate},
one at time, with the CMB constraint. We did not include the
{\it XMM} observations of the Milky Way because they are discontinous
and this makes the mass-averaging procedure problematic.
The results are illustrated in
Fig. \ref{fig:best-constraints}. We can roughly say that 
for small majoron masses ($\beta\sim 1$), we should resort to x-ray observations
to probe the region $R\gtrsim10^{-6}$ , while
we should use the CMB for $R\lesssim10^{-6}$. 
For large neutrino masses, x-ray observations are better
when $R\gtrsim 10^{-8}$, while CMB is more informative
for $R\lesssim 10^{-10}$.

\section{Particle physics}
\label{sec:particle-physics}

We now turn to the particle physics of our decaying dark matter
scenario. Although many attractive options are
open~\cite{Nunokawa:2007qh} possibly the most popular scheme for
generating neutrino masses is the seesaw.  The simplest type I seesaw
model has no induced majoron radiative decays.  For this reason we
consider the full seesaw model, which contains a Higgs boson triplet
coupling to the lepton doublets~\cite{schechter:1980gr}.

In addition to the SM fields one has three electroweak gauge
singlet right-handed neutrinos, $\nu^c_{L_i}$, a complex $SU(2)_L$
scalar triplet $\Delta$, with hypercharge $1$ and lepton number $-2$ ,
and a scalar singlet $\sigma$, with lepton number $2$.  We will denote
the scalar $SU(2)$ doublet as $\phi$. The Yukawa Lagrangian is given
by
\begin{eqnarray}
\label{eq:LY}
\mathcal{L}_Y&=& Y_u {Q}^T_L \phi u_L^c+ Y_d {Q}^T_L \phi^* d_L^c + Y_e {L}^T_L \phi^* e_L^c \nn\\
&+& Y_\nu {L}^T_L \phi \nu_L^c + {Y}_L  L^T_L \Delta L_L + \frac{Y_R}{2} \nu_L^c \nu_L^c  \sigma + H.c.
\end{eqnarray}
In order to extract the relevant couplings of the majoron that are
responsible for the decays in Eq.~(\ref{eq:gammanu-limit}), we review
here the main steps of the procedure developed
in~\cite{schechter:1982cv}, using the basic two-component Weyl
description of neutrinos as in \cite{schechter:1980gr}.

Using the invariance of the scalar potential under the hypercharge
$U(1)_Y$ and lepton number $U(1)_L$ symmetries and assuming that these
are broken spontaneously by the vacuum configuration, one finds, from
Noether's theorem, the full structure of the mass matrix of the
imaginary neutral component of the scalars given in terms of their
vacuum expectation values (vevs) as~\cite{schechter:1982cv}
\begin{eqnarray}
\label{eq:psmass}
M^{I 2}&=&
C\,\left(
\begin{array}{ccc}
4 \frac{v_1^2}{v_2^2}  & -2\frac{v_1^2}{v_3 v_2}     &  2 \frac{v_1}{v_2}  \\
-2\frac{v_1^2}{v_3 v_2}     &   \frac{v_1^2}{v_3^2}&   - \frac{v_1}{v_3} \\
2 \frac{v_1}{v_2}   & - \frac{v_1}{v_3}   &   1
\end{array}
\right)\,,
\end{eqnarray}
with $C={\partial^2 V}/{\partial^2 \sigma^I}$ and $v_1,v_2,v_3$ are
the vevs of the singlet, the doublet, and the triplet, respectively. 
One sees that $M^{I 2}$ has a non zero eigenvalue, $m^2_A= \Tr M^{I2}$
and two null eigenvalues. These correspond to the Goldstone bosons
eaten by the $Z$ gauge boson and, as expected, to the physical
Nambu-Goldstone boson associated with the breaking of $U(1)_L$, the
majoron $J$. The parameters of the scalar potential of the model can
be chosen so that the pattern of vevs obtained by minimization
respects the so--called (type II) seesaw form,
namely~\cite{schechter:1982cv}
$$v_3 \ll v_2 \ll v_1$$
In particular, since the smallness of the triplet vev arises through
the vev seesaw relation one can show that it is not spoiled by
one-loop radiative corrections.

The resulting profile of the majoron, $J$, following from
Eq.~(\ref{eq:psmass}) takes a very simple form in this seesaw
approximation, namely~\cite{schechter:1982cv}
\begin{equation}
\label{eq:proJ}
 J\simeq    -\frac{ 2 v_3^2}{v_1 v_2} \,\phi^{0I}  +   \frac{ v_3}{ v_1} \,\Delta^{0I} + \sigma^{0I}  \,.
 \end{equation}

 In the presence of the gravitationally induced terms that give mass
 to the majoron, the mass matrix of Eq.~(\ref{eq:psmass}) is slightly
 modified, but these effects are sub-leading and negligible. 

 We are now ready to determine the coupling of the majoron with the
 light neutrinos. From Eq.~(\ref{eq:LY}) one obtains the full neutrino
 Majorana mass matrix as
\begin{eqnarray}
\label{eq:Mnu0}
M^\nu&=&\frac{1}{2}
\left(
\begin{array}{cc}
{Y}_L v_3  &  Y_\nu v_2   \\
  Y_\nu v_2 &   Y_R v_1  
\end{array}
\right)\,.
\end{eqnarray}
so that the effective light neutrino Majorana mass matrix is given
by~\cite{schechter:1982cv}:
\begin{eqnarray}
\label{eq:Mnu}
M^\nu_{LL}&=& \frac{1}{2} \left( {Y}_L v_3 - Y_\nu ^T  Y_R^{-1} Y_\nu \frac{v_2^2}{v_1}\right)\,.
\end{eqnarray}
The coupling $g_{J\nu\nu}$ of the majoron to the neutrinos can also be
obtained using Noether's theorem according to the
procedure described in Ref.~\cite{schechter:1982cv}.  In this way
  one finds that the majoron couples to the mass eigenstate neutrinos
  proportionally to their masses,
\begin{equation}
g^\nu_{J r s  } = - \frac{m^\nu_r \delta_{rs}}{2 v_1}\,.
\end{equation}
where $v_1$ describes the scale at which the global lepton number
symmetry breaks, typically $10^6-10^9$ GeV (see below).

The decay width $\Gamma_{J \nu \nu} $ is given by
\begin{equation}
\Gamma_{J \nu \nu} = \frac{m_J}{32 \pi} \frac{\Sigma_r (m^\nu_r)^2}{4 v_1^2}\,,
\end{equation}

Let's now turn to $g_{J\gamma\gamma}$.
From the Yukawa Lagrangian of Eq.~(\ref{eq:LY}) and from
Eq.~(\ref{eq:proJ}) we have that the majoron interacts with the charged
fermions through

\begin{equation}
\label{eq:intJ}
 -\frac{ 2 v_3^2}{v_2 v_1} Y_f (-2 T_{3f})\bar{f} \gamma_5 f\, J = -\frac{ 2 v_3^2}{v_2^2 v_1} m_f (-2 T_{3f}) \bar{f} \gamma_5 f\,J\,,
\end{equation}
where $T_{3f}$ is the weak isospin and we have assumed that the
charged fermion mass matrices are diagonal. The interaction term of
Eq.~(\ref{eq:intJ}) gives rise to the interaction term with photons
given in Eq.~(\ref{eq:Jgg}), with an effective coupling given by
\begin{equation}
g_{J\gamma\gamma} =\frac{\alpha}{2 \pi} \Sigma_f \,N_f\,(-2 T_{3f})\,Q_f^2 \left(1+\frac{1}{12}\frac{m_J^2}{m_f^2}\right)\frac{ 2 v_3^2}{v_2^2 v_1}\,.
\end{equation}
where one notices the cancellation of the ``anomalous-like'' contribution $\Sigma_f \,N_f\,(-2 T_{3f})\,Q_f^2$.
As a result we have
\begin{equation}
\label{eq:gg}
\Gamma_{J \gamma \gamma}= \frac{\alpha^2}{64 \pi^3} \frac{m_J^3}{\tilde{\Lambda}_\gamma^2}\,,
 \end{equation}
with
$$\tilde{\Lambda}_\gamma = \frac{1}{\Sigma_f \,N_f\,(-2 T_{3f}) Q_f^2\frac{1}{12}~\frac{m_J^2}{m_f^2} }\frac{v_2^2 v_1}{v_3^2}\,,
$$
where $Q_f$ and $N_f$ are the electric charge of $f$ and its
colour factor, respectively.

\begin{center}
\begin{figure}[h!]                                      
\includegraphics[clip,width=0.5\linewidth]{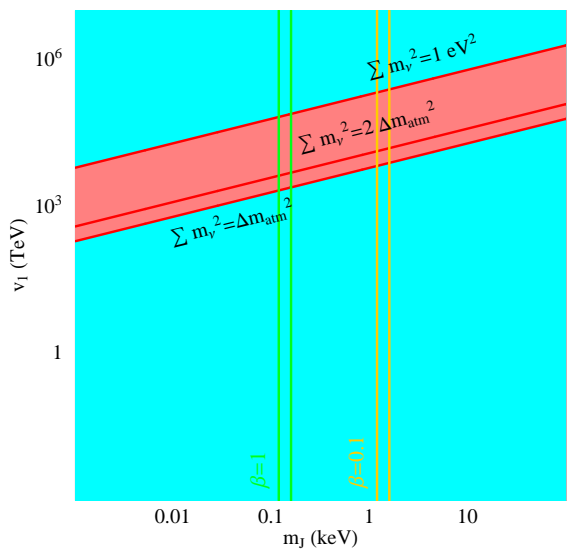}
\caption{The strip indicates the region in the $v_1-m_J$ plane allowed
  by current neutrino oscillation~\cite{Maltoni:2004ei} and
  cosmological data \cite{Dunkley:2008ie}, assuming the maximal $J \to
  \nu \nu$ decay rate. The vertical lines delimit the mass values
  required by the CMB observations, for different values of $\beta$.}
\label{fig:region-mJ-v1.eps}
\end{figure}
\end{center}

Fig.~\ref{fig:region-mJ-v1.eps} shows how the currently allowed range
of neutrino masses selects an allowed strip in the plane $v_1-m_J$
consistent with neutrino oscillation data~\cite{Maltoni:2004ei} and
with the cosmological bounds on neutrino mass \cite{Dunkley:2008ie},
assuming that the CMB bound~(\ref{eq:gammanu-limit}) on the $J \to \nu
\nu$ decay rate is saturated.  

 The lower lines correspond to the cases of normal and inverse
  hierarchical neutrino masses, while the top line holds when
  the three neutrinos are (quasi)-degenerate.
  The vertical bands in the figure indicate the mass region of
  Eq.~\ref{eq:mj-constraint} singled out by the CMB observations, for
  two different values of $\beta$.

\begin{center}
 \begin{figure}[h!]
  \includegraphics[clip,width=0.5\linewidth]{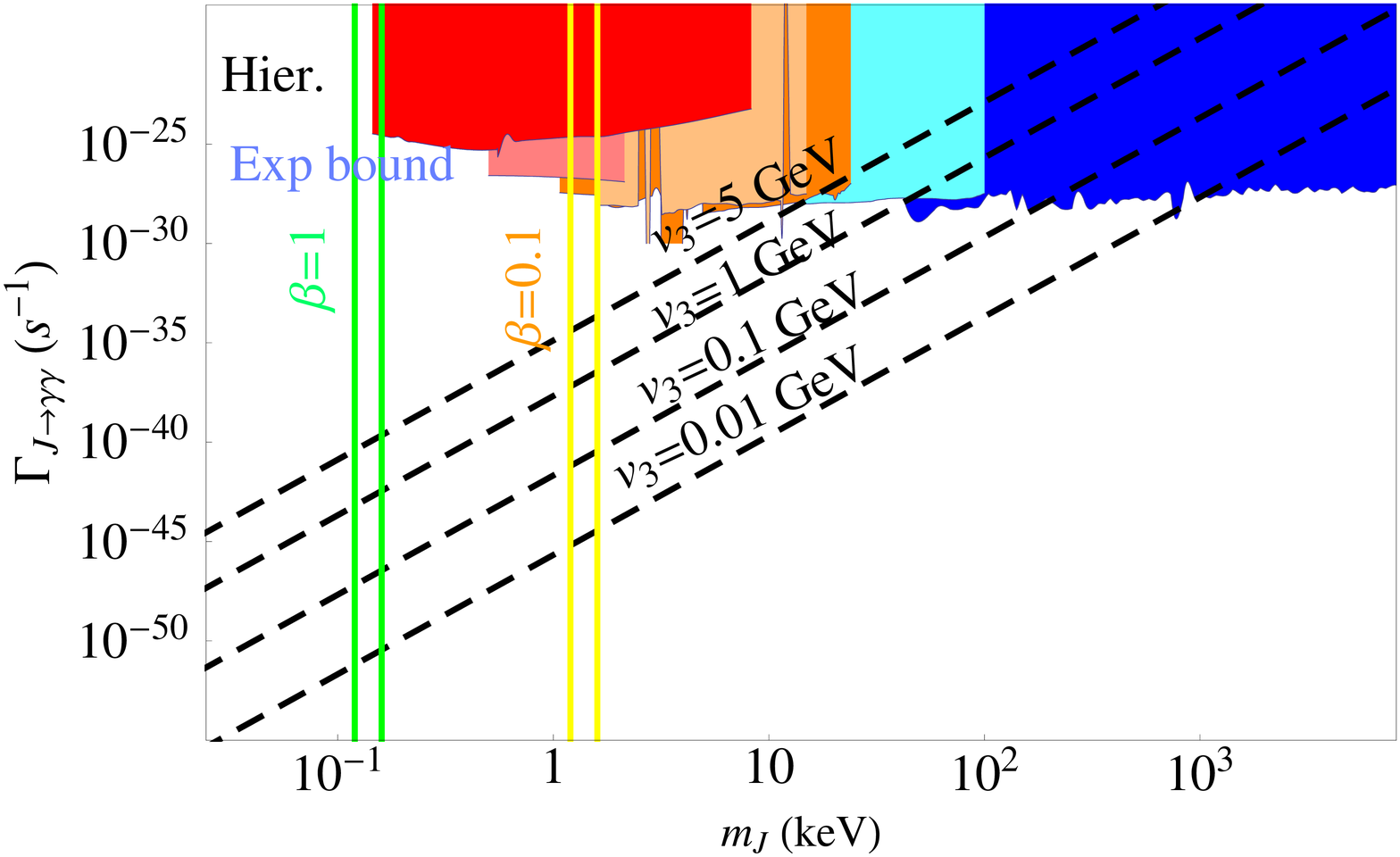}
  \includegraphics[clip,width=0.5\linewidth]{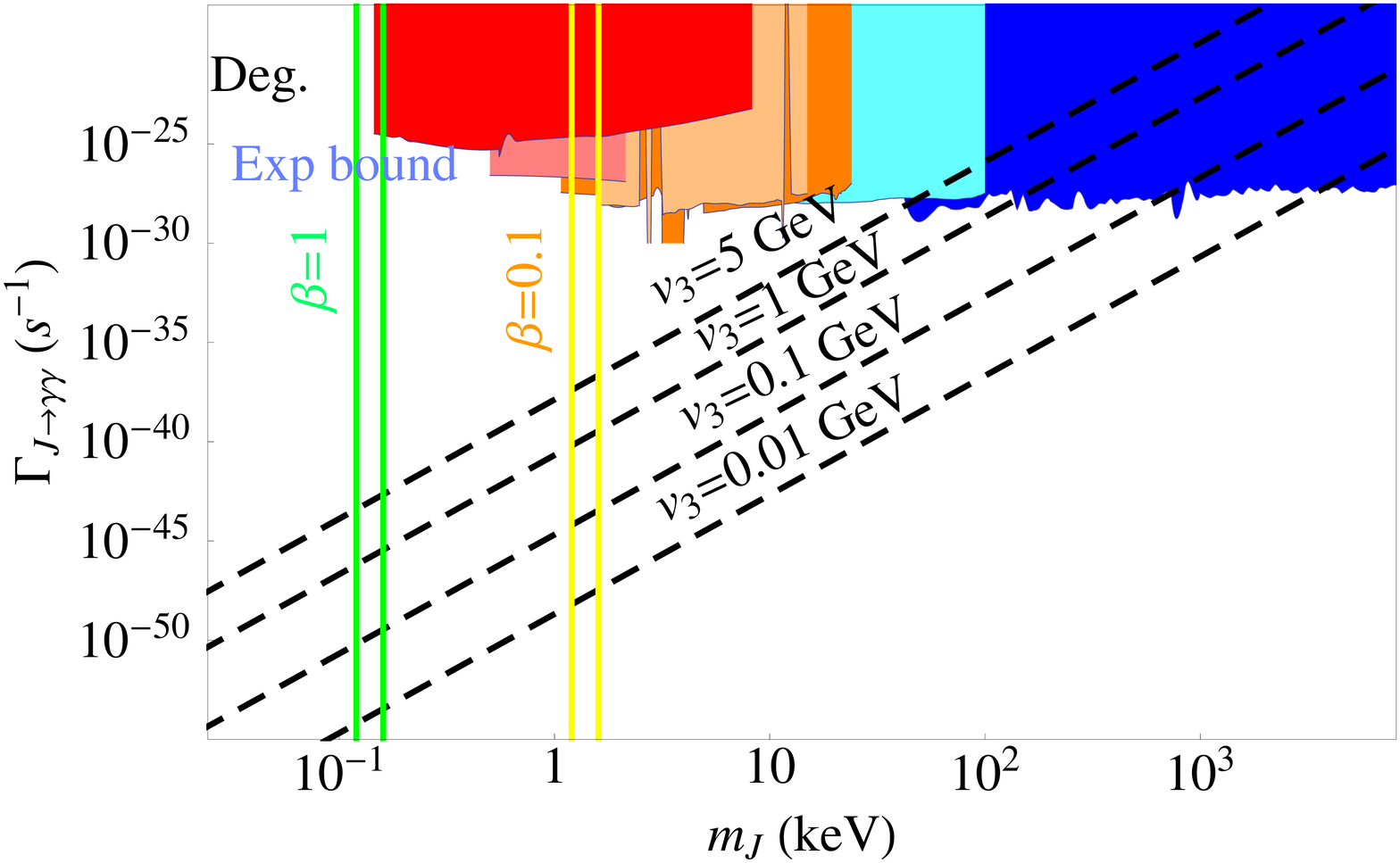}
  \caption{ Majoron decay rate to photons as a function of the
      majoron mass $m_J$, for different values of the triplet vev,
      $v_3$. We assume the invisible decay bound to be saturated. The
      top and bottom panels refer to hierarchical and degenerate
      neutrino mass spectra, respectively.  The shaded regions are
      excluded by observations as described in Sec.~\ref{sec:x-rays}.
      The vertical lines are the same as in
      Fig. \ref{fig:region-mJ-v1.eps}.}           
  \label{fig:decayJgg.eps}
 \end{figure}
\end{center}

We also note from Eq.~(\ref{eq:gg}) that, for a fixed value of the
majoron mass and the lepton number symmetry breaking scale $v_1$, the
two-photon decay rate only depends on the vev of the triplet and on
the sum of the squared masses of the neutrinos, namely on the two
possible scenarios in the neutrino sector, hierarchical or
degenerate. For a given scenario the decay is then fixed only by
$v_3$, as it can been seen in Fig.~\ref{fig:decayJgg.eps}, where the
top panel corresponds to the hierarchical case while the bottom one
holds for the quasi-degenerate spectrum.
The diagonal lines in Fig.~\ref{fig:decayJgg.eps} give the
dependence of $\Gamma_{J\gamma\gamma}$ on $m_J$, for different
values of $v_3$. As it can also be seen from Eq. (\ref{eq:gg}), the largest values of
$v_3$ correspond to the largest radiative decay rates. 

One sees that in both scenarios small $m_J$ and $v_3$ values
    lead to decay rates well below the observational bounds.
    However, for large values of $v_3$, say, $v_3=5$~GeV, roughly
    corresponding to the maximum compatible with precision
    measurements of electroweak parameters~\cite{Yao:2006px}, the
    radiative rates fall within the sensitivities of the Milky Way
    observations displayed in Fig.~\ref{fig:decayrate}, and would be
    thereby observationally excluded. For lower masses the
    observational sensitivities would need to be improved by about 20
    orders of magnitude requiring completely new techniques from what
    is available today. The small radiative majoron decay rates would be
    avoided in models where the anomaly does not cancel due to the
    presence of extra fermions. We mention also in this case the
    possibility of further enhancement due to cumulative effects as those that
    might arise, for example, in higher dimensions.

\section{Summary}

We have investigated the production of x-ray photons in the
late-decaying dark matter scenario, and quantified the sensitivity of
current observations to such a mono-energetic emission line. In
particular, we have studied the constraints from the diffuse x-ray
observations, as well as by considering the fluxes generated by dark
matter dominated objects. These observations provide a probe of
  radiative dark matter decays and can be used as an ``indirect
  detection'' of the LDDM majoron scenario.

  We have illustrated this explicitly for the case where neutrinos get
  mass \textsl{a la seesaw}, where the majoron couples to photons
  through its Higgs triplet admixture. Alternative particle physics
  realizations of the LDDM scenario can be envisaged, an issue which
  will be taken up elsewhere.
  Let us also mention that Majoron dark matter decays can be possibly
  probed in the future through 21-cm observations (see
  Ref.~\cite{Valdes:2007cu} for an application to other DM
  candidates).

\paragraph{Acknowledgements} We thank Antonio Palazzo and Alexei Boyarsky
for discussions. This work was supported by MEC 
grant FPA2005-01269, by EC Contracts RTN network MRTN-CT-2004-503369
and ILIAS/N6 RII3-CT-2004-506222. 
The Dark Cosmology Centre is funded by the Danish National Research Foundation.
ML is currently supported by INFN.


\section*{References}

\begin{thebibliography}{10}

\bibitem{fukuda:1998mi}
Super-Kamiokande collaboration, Y.~Fukuda {\em et~al.},
\newblock Phys. Rev. Lett. {\bf 81}, 1562 (1998), [hep-ex/9807003].

\bibitem{ahmad:2002jz}
SNO collaboration, Q.~R. Ahmad {\em et~al.},
\newblock Phys. Rev. Lett. {\bf 89}, 011301 (2002), [nucl-ex/0204008].

\bibitem{eguchi:2002dm}
KamLAND collaboration, K.~Eguchi {\em et~al.},
\newblock Phys. Rev. Lett. {\bf 90}, 021802 (2003), [hep-ex/0212021].

\bibitem{kl:2008ee}
KamLAND collaboration, S.~Abe {\em et~al.},
\newblock 0801.4589.

\bibitem{Collaboration:2007zza}
MINOS collaboration,
\newblock arXiv:0708.1495 [hep-ex].

\bibitem{Maltoni:2004ei}
M.~Maltoni, T.~Schwetz, M.~A. Tortola and J.~W.~F. Valle,
\newblock New J. Phys. {\bf 6}, 122 (2004), 
 \newblock arXiv version 6 in hep-ph/0405172 provides updated
  neutrino oscillation results and references to previous works.

\bibitem{Drexlin:2005zt}
KATRIN  collaboration, G.~Drexlin,
\newblock Nucl. Phys. Proc. Suppl. {\bf 145}, 263 (2005).

\bibitem{Avignone:2007fu}
I.~Avignone, Frank~T., S.~R. Elliott and J.~Engel,
\newblock 0708.1033.
a brief summary of the phenomenology of double beta decay 
can be found in: M.~Hirsch,
arXiv:hep-ph/0609146.

\bibitem{Lesgourgues:2006nd}
S.~Hannestad,
  New J.\ Phys.\  {\bf 6}, 108 (2004)
  [arXiv:hep-ph/0404239].
J.~Lesgourgues and S.~Pastor,
\newblock Phys. Rept. {\bf 429}, 307 (2006), [astro-ph/0603494].

\bibitem{Chikashige:1980ui} Y.~Chikashige, R.~N. Mohapatra and
  R.~D. Peccei, \newblock Phys. Lett. {\bf B98},  26 (1981). 

\bibitem{schechter:1982cv}
J.~Schechter and J.~W.~F. Valle,
\newblock Phys. Rev. {\bf D25}, 774 (1982).

\bibitem{Coleman:1988tj}
S.~R. Coleman,
\newblock Nucl. Phys. {\bf B310}, 643 (1988);
  R.~Holman, S.~D.~H.~Hsu, T.~W.~Kephart, E.~W.~Kolb, R.~Watkins and L.~M.~Widrow,
  Phys.\ Lett.\  B {\bf 282} (1992) 132
  [arXiv:hep-ph/9203206];
 E.~K.~Akhmedov, Z.~G.~Berezhiani and G.~Senjanovic,
  Phys.\ Rev.\ Lett.\  {\bf 69} (1992) 3013
  [arXiv:hep-ph/9205230].

\bibitem{Berezinsky:1993fm}
  V.~Berezinsky and J.~W.~F.~Valle,
  Phys.\ Lett.\  B {\bf 318} 360 (1993) 
  [arXiv:hep-ph/9309214].

\bibitem{Spergel:2006}
WMAP collaboration, D.~N. Spergel {\em et~al.},
\newblock Astrophys. J. Suppl. {\bf 170}, 377 (2007), [astro-ph/0603449].

\bibitem{Lattanzi:2007ux}
M.~Lattanzi and J.~W.~F. Valle,
\newblock Phys. Rev. Lett. {\bf 99}, 121301 (2007), [arXiv:0705.2406
  [astro-ph]].

\bibitem{Berezinsky:1996pb}
V.~Berezinsky, A.~S.~Joshipura and J.~W.~F.~Valle,
  Phys.\ Rev.\  D {\bf 57}, 147 (1998) 
  [arXiv:hep-ph/9608307].

\bibitem{Ichiki:2004vi}
  K.~Ichiki, M.~Oguri and K.~Takahashi,
  Phys.\ Rev.\ Lett.\  {\bf 93}, 071302 (2004) 
  [arXiv:astro-ph/0403164].

\bibitem{Gong:2008gi}
  Y.~Gong and X.~Chen,
  arXiv:0802.2296 [astro-ph].
 
\bibitem{Zhang:2007zzh}
  L.~Zhang, X.~Chen, M.~Kamionkowski, Z.~g.~Si and Z.~Zheng,
  Phys.\ Rev.\  D {\bf 76}, 061301 (2007)
  [arXiv:0704.2444 [astro-ph]].
  
\bibitem{Chen:2003gz}
  X.~L.~Chen and M.~Kamionkowski,
  Phys.\ Rev.\  D {\bf 70}, (2004) 043502 
  [astro-ph/0310473].
  	
\bibitem{Gendreau:1995}  
  K.~C.~Gendreau et al.,
  Publ.\ Astron.\ Soc.\ Jap.\  {\bf 47}, L5 (1995).  

\bibitem{Gruber:1999yr}
  D.~E.~Gruber, J.~L.~Matteson, L.~E.~Peterson and G.~V.~Jung,
  Astrophys.\ J.\  {\bf 520}, 124 (1999).
  [arXiv:astro-ph/9903492].
  
\bibitem{Kawasaki:1997ah}
  M.~Kawasaki and T.~Yanagida,
  Phys.\ Lett.\  B {\bf 399}, 45 (1997)
  [arXiv:hep-ph/9701346].
  
\bibitem{Chen:1995jc}
  L.~W.~Chen, A.~C.~Fabian and K.~C.~Gendreau,
  Mon.\ Not.\ R.\ Astron.\ Soc.\ {\bf 285}, 449 (1997)
  [arXiv:astro-ph/9511089].
  
\bibitem{Boyarsky:2005us}
  A.~Boyarsky, A.~Neronov, O.~Ruchayskiy and M.~Shaposhnikov,
  Mon.\ Not.\ Roy.\ Astron.\ Soc.\  {\bf 370}, 213 (2006)
  [arXiv:astro-ph/0512509].

\bibitem{Boyarsky:2006fg}
  A.~Boyarsky, A.~Neronov, O.~Ruchayskiy, M.~Shaposhnikov and I.~Tkachev,
  Phys.\ Rev.\ Lett.\  {\bf 97}, 261302 (2006)
  [arXiv:astro-ph/0603660].
  
\bibitem{Boyarsky:2006hr}
  A.~Boyarsky, J.~W.~den Herder, A.~Neronov and O.~Ruchayskiy,
  Astropart.\ Phys.\  {\bf 28}, 303 (2007)
  [arXiv:astro-ph/0612219].

\bibitem{POG9}
C.~X. ray Centre, {\bf 644}, L33 (2006)
\newblock http://cxc.harvard.edu/proposers/POG/html 

\bibitem{Riemer:2006b}
S.~Riemer-Sorensen, S.~H. Hansen, K.~Pedersen and H.~Dahle,
\newblock Phys. Rev. D {\bf 76}, 043524 (2007).

\bibitem{Boyarsky:2006kc}
  A.~Boyarsky, O.~Ruchayskiy and M.~Markevitch,
  Astrophys.\ J.\  {\bf 673}, 752 (2008)
  [arXiv:astro-ph/0611168].

\bibitem{Fruscione:2006}
A.~Fruscione et al. SPIE Proc. 6270, 62701V, D.R. Silvia \& R.E. Doxsey, eds.
(2006).\\
\newblock http://cxc.harvard.edu/ciao/

\bibitem{Navarro:1996gj}
J.~F. Navarro, C.~S. Frenk and S.~D.~M. White,
\newblock Astrophys. J. {\bf 490}, 493 (1997), [astro-ph/9611107].

\bibitem{Moore:1999gc}B.~Moore, T.~Quinn, F.~Governato, J.~Stadel and G.~Lake,\newblock Mon.\ Not.\ Roy.\ Astron.\ Soc.\  {\bf 310}, 1147 (1999)

\bibitem{Bentz:2006qc}
M.~C. Bentz, B.~M. Peterson, R.~W. Pogge, M.~Vestergaard and C.~A. Onken,
\newblock Astrophys. J. {\bf 644}, 133 (2006), [astro-ph/0602412].

\bibitem{Boyarsky:2006ag}
  A.~Boyarsky, J.~Nevalainen and O.~Ruchayskiy,
  Astron.\ Astrophys.\  {\bf 471}, 51 (2007)
  [arXiv:astro-ph/0610961].
  
\bibitem{Watson:2006qb}
  C.~R.~Watson, J.~F.~Beacom, H.~Yuksel and T.~P.~Walker,
  Phys.\ Rev.\  D {\bf 74}, 033009 (2006)
  [arXiv:astro-ph/0605424].

\bibitem{Boyarsky:2007ay}  A.~Boyarsky, D.~Iakubovskyi, O.~Ruchayskiy and V.~Savchenko,  
  arXiv:0709.2301 [astro-ph].  
  
\bibitem{Yuksel:2007xh}
  H.~Yuksel, J.~F.~Beacom and C.~R.~Watson,
  arXiv:0706.4084 [astro-ph].

\bibitem{Boyarsky:2007ge}
A.~Boyarsky, D.~Malyshev, A.~Neronov and O.~Ruchayskiy,
\newblock arXiv:0710.4922 [astro-ph].

\bibitem{Nunokawa:2007qh}
H.~Nunokawa, S.~J. Parke and J.~W.~F. Valle,
\newblock Prog. Part. Nucl. Phys. {\bf 60}, 338 (2008), [arXiv:0710.0554
  [hep-ph]].

\bibitem{schechter:1980gr}
J.~Schechter and J.~W.~F. Valle,
\newblock Phys. Rev. {\bf D22}, 2227 (1980).

\bibitem{Dunkley:2008ie}
  J.~Dunkley {\it et al.}  [WMAP Collaboration],
  arXiv:0803.0586 [astro-ph].

\bibitem{Yao:2006px}
Particle Data Group, W.~M. Yao {\em et~al.},
\newblock J. Phys. {\bf G33}, 1 (2006).

\bibitem{Valdes:2007cu}
  M.~Valdes, A.~Ferrara, M.~Mapelli and E.~Ripamonti,
  Mon.\ Not.\ Roy.\ Astron.\ Soc.\  {\bf 377}, 245 (2007)
  [arXiv:astro-ph/0701301].

\end{thebibliography}
\end{document}